\documentclass[
aps,
reprint,
superscriptaddress,
amssymb,
amsmath
]{revtex4-1}

\usepackage{graphicx}
\usepackage{dcolumn}
\usepackage{bm}
\usepackage{bm, color}
\usepackage{bbm}
\usepackage{lineno}
\usepackage{braket}

\begin{document}

\title{Identification of chirality of chiral multifold fermions in anti-crystals}

\author{Yan Sun}
\email{ysun@cpfs.mpg.de}
\affiliation{Max Planck Institute for Chemical Physics of Solids, 01187 Dresden, Germany}
\author{Qiunan Xu}
\affiliation{Max Planck Institute for Chemical Physics of Solids, 01187 Dresden, Germany}
\author{Yang Zhang}
\affiliation{Department of Physics, Massachusetts Institute of Technology, Cambridge, Massachusetts 02139, USA}
\author{Congcong Le}
\affiliation{Max Planck Institute for Chemical Physics of Solids, 01187 Dresden, Germany}
\author{Claudia Felser}
\affiliation{Max Planck Institute for Chemical Physics of Solids, 01187 Dresden, Germany}
\affiliation{Center for Nanoscale Systems, Faculty of Arts and Sciences, Harvard University, 11 Oxford Street, LISE 308 Cambridge, Massachusetts 02138, USA}

\begin{abstract}
The chirality of chiral multifold fermions in reciprocal
space is related to the chirality of the crystal lattice
structure in real space. In this work, we propose a
strategy to detect and identify opposite-chirality
multifold fermions in nonmagnetic systems
by means of second-order optical transports.
The chiral crystals are related by an inversion operation and
cannot overlap with each other by any experimental operation,
and the chiral multifold fermions in the crystals host opposite
chiralities for a given $k$-point. A change of chirality is indicated by a sign change of
the second-order charge current dominated by chiral fermions.
This strategy
is effective to study the
relationship between chiralities in reciprocal and real spaces
by utilizing bulk transport.
\end{abstract}

\maketitle

Multifold massless fermions with nonzero topological
charge are attracting increasing attention in the field of topological materials.
In contrast to four-fold degenerate Dirac and doubly degenerate
Weyl fermions, which have directly
analogous fundamental particles in high-energy physics,
multifold fermions do not follow the Poincar\'e symmetry
in high-energy physics; rather, they follow the crystal symmetry
and, therefore, do not have counterparts among real particles.
~\cite{manes2012existence,bradlyn2016beyond}.
More interestingly, the multifold fermions in chiral crystals
can host nonzero topological charges with a Chern number greater
than 1, which results in both nontrivial topological
surface states and exotic bulk transport properties.

Since multifold fermions are located at high-symmetry points, 
they guarantee long surface Fermi arcs spanning the 
entire Brillouin zone (BZ)
~\cite{chang2017unconventional,tang2017multiple,chang2018,zhang2018double}. 
In addition to the huge separation of opposite topological
charges in momentum space, the absence of mirror symmetry 
leads to their large separation in energy space, providing
an ideal platform for the study of the quantized circular 
photogalvanic effect (CPGE)~\cite{chang2017unconventional,chang2018,flicker2018chiral,de2017quantized}. 
Soon after theoretical prediction,
the long Fermi arcs, high-order degenerated band crossings,
and quantized CPGE were observed in the predicted
chiral crystals by 
ARPES~\cite{sanchez2019topological,schroter2019chiral,rao2019observation,takane2019observation, schroter2019observation} 
and optical measurements~\cite{rees2019quantized}.

\begin{figure}[htbp]
\begin{center}
\includegraphics[width=0.5\textwidth]{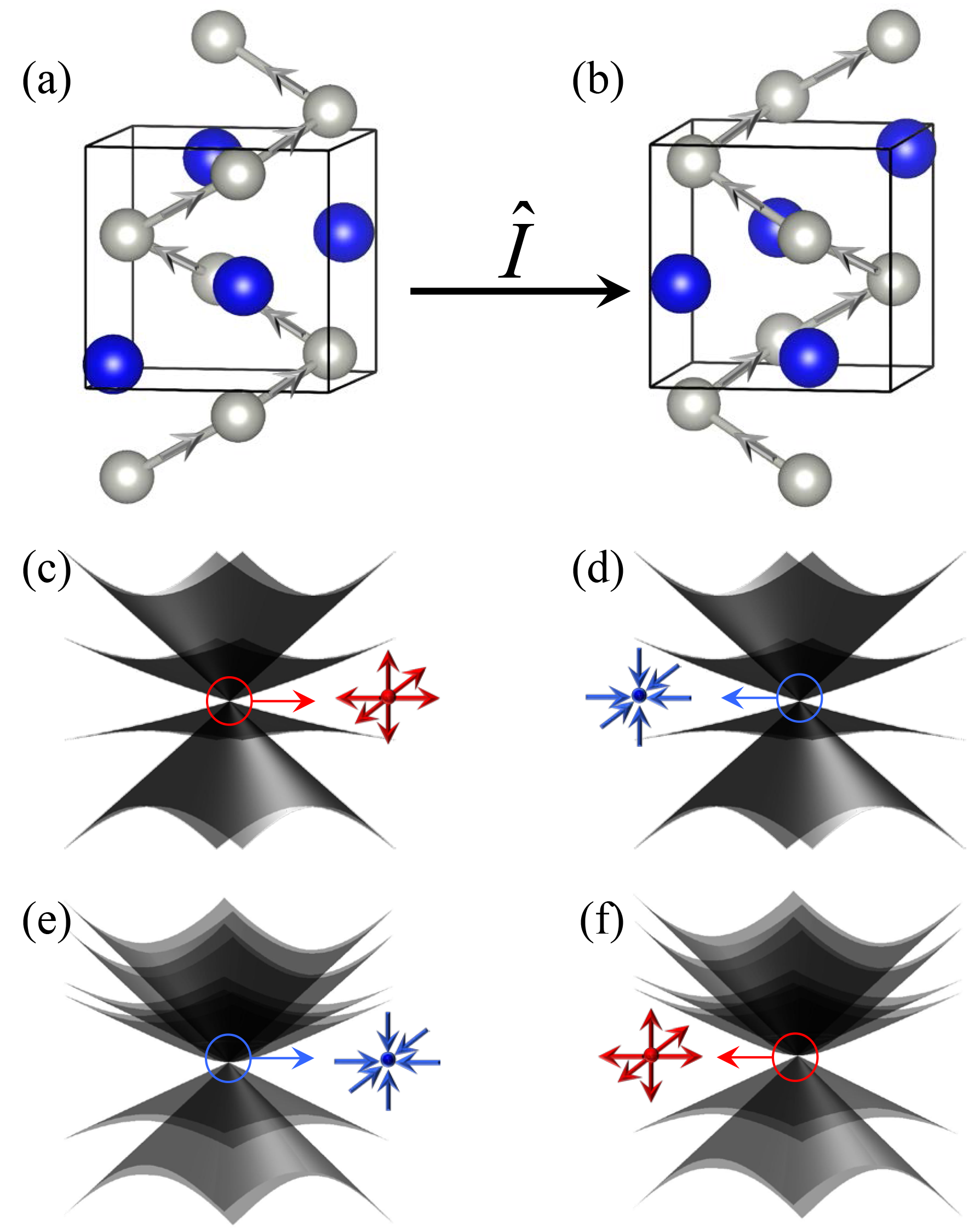}
\end{center}
\caption{
        Inversion of chirality in a chiral crystal structure
        leads to the inversion of chirality for chiral multifold fermions.
        (a, b) Crystal lattice structures with opposite chiralities
        of semimetals in the space group P213 with chiral multifold fermions,
        taking RhSi as an example.
        (c, d) Four-fold degenerate fermion with spin-3/2 excitation at
        the $\Gamma$ point.
        (e, f) Double three-fold degenerate fermions with spin-1 excitation at
        the $R$ point. The local effective Hamiltonians around crossing points are
        $H=-\vec{k}\cdot\vec{S}_{4\times4}$,
        $H=\vec{k}\cdot\vec{S}_{4\times4}$,
        $H=-\vec{k}\cdot\vec{S}_{3\times3}$, and
        $H=\vec{k}\cdot\vec{S}_{3\times3}$ for (c-f), respectively.
}
\label{crystal}
\end{figure}

Recently, via ARPES measurements, a sign change of the Fermi
velocity of surface Fermi arcs was observed in semimetals having chiral
multifold fermions and crystals of opposite chiralities,
implying a deep relationship between the chiral lattices in real
space and the chiral fermions in momentum space
~\cite{schroter2019observation}.
This relationship offers a new degree of freedom to tune the chiral fermions
and corresponding physical properties. Hence, it is important
to understand the relationship of chirality between
real and reciprocal spaces as well as the resulting phenomena.
In the present study, based on a symmetry analysis and numerical
calculations, we provide a strategy for detecting
the interplay between chiral fermions and chiral crystals
from the transport perspective.
Since the crystal structures of the same compound with
opposite chiralities are related by an inversion operation,
the relationship between chiral fermions
and chiral crystals can be detected by means of
nonlinear optical and electrical transports.

Here, we focus on materials in the space group 
$P2_13$ because, to date, all the experimentally verified materials with 
chiral multifold fermions belong to this space group. 
For a crystal with the space group 
$P2_13$, owing to the glide two-fold rotation
symmetry ($s_{2x}=\left\{ c_{2x}|(\frac{1}{2}\frac{1}{2}0)\right\} $,
$s_{2y}=\left\{ c_{2y}|(0\frac{1}{2}\frac{1}{2})\right\} $, and 
$s_{2z}=\left\{ c_{2z}|(\frac{1}{2}0\frac{1}{2})\right\}$),
corresponding glide mirror operations can be obtained
through a simple inversion operation. Therefore, chiral crystals
with opposites chiralities in the space group $P2_13$ are connected 
by a simple inversion operation, as shown in Fig. 1(a,b).
In any other crystal structure without $c_2$ or glide $c_2$ 
rotation symmetry, a mirror operation can be also equivalent 
to an inversion operation in combination with an experimental 
$c_2$ rotation operation
because samples can be rotated by any angle in experiments.

In compounds with the space group $P2_13$, the chiral multifold
fermions located at $\Gamma$ and $R$ have four-fold and
six-fold degeneracy, respectively
~\cite{manes2012existence,bradlyn2016beyond,chang2017unconventional,tang2017multiple,chang2018,zhang2018double}. 
Similar to Weyl fermions,  
these two types of chiral multifold fermions can be 
described in a unified form using the Hamiltonian $H=\chi\vec{k}\cdot\vec{S}$, with
$\chi=\pm1$~\cite{bradlyn2016beyond} and the pseudospin matrix following
$[S_{i},S_{j}]=i\epsilon_{ijk}S_{k}$.
The four-fold degeneracy at $\Gamma$ points is a 
spin-3/2 excitation with a $4\times4$ pseudospin matrix, and the
topological charge has a Chern number of $\pm4$ when the two lower bands
are occupied. The six-fold degeneracy at the $R$ point is constructed
by double spin-1 Weyl fermions
and a doubly degenerate quadratic band; hence, the
 $3\times3$ pseudospin matrix can be used to describe this six-fold degeneracy, as shown in
 Fig. 1(e). Since the spin-1 Weyl fermions have a Chern number of
 $\pm2$, the corresponding topological charge for double spin-1 Weyl fermions
 is $\pm4$, making the whole system follow a ``no-go theorem.''
 For both spin-1 and spin-3/2 excitation, the sign of the topological
 charge is dependent on the sign of the prefactor $\chi$. Under an inversion
 operation, $H(\vec{k})$ is changed to $H(-\vec{k})$. While keeping
 the form of energy dispersion, chirality changes the sign, as
 is evident from a comparison of Fig. 1 (c-f).

Though the chirality of a multifold fermion is reversed by the 
inversion operation, time-reversal symmetry makes the 
net Berry phase zero in the entire
BZ. Therefore, the anomalous 
Hall effect cannot be used to probe the change of chirality. 
However, the second-order responses are odd with respect to 
inversion, which provides the possibility to detect the
change in chirality of multifold fermions of semimetals in anti-crystals.

\begin{figure}[htbp]
\begin{center}
\includegraphics[width=0.5\textwidth]{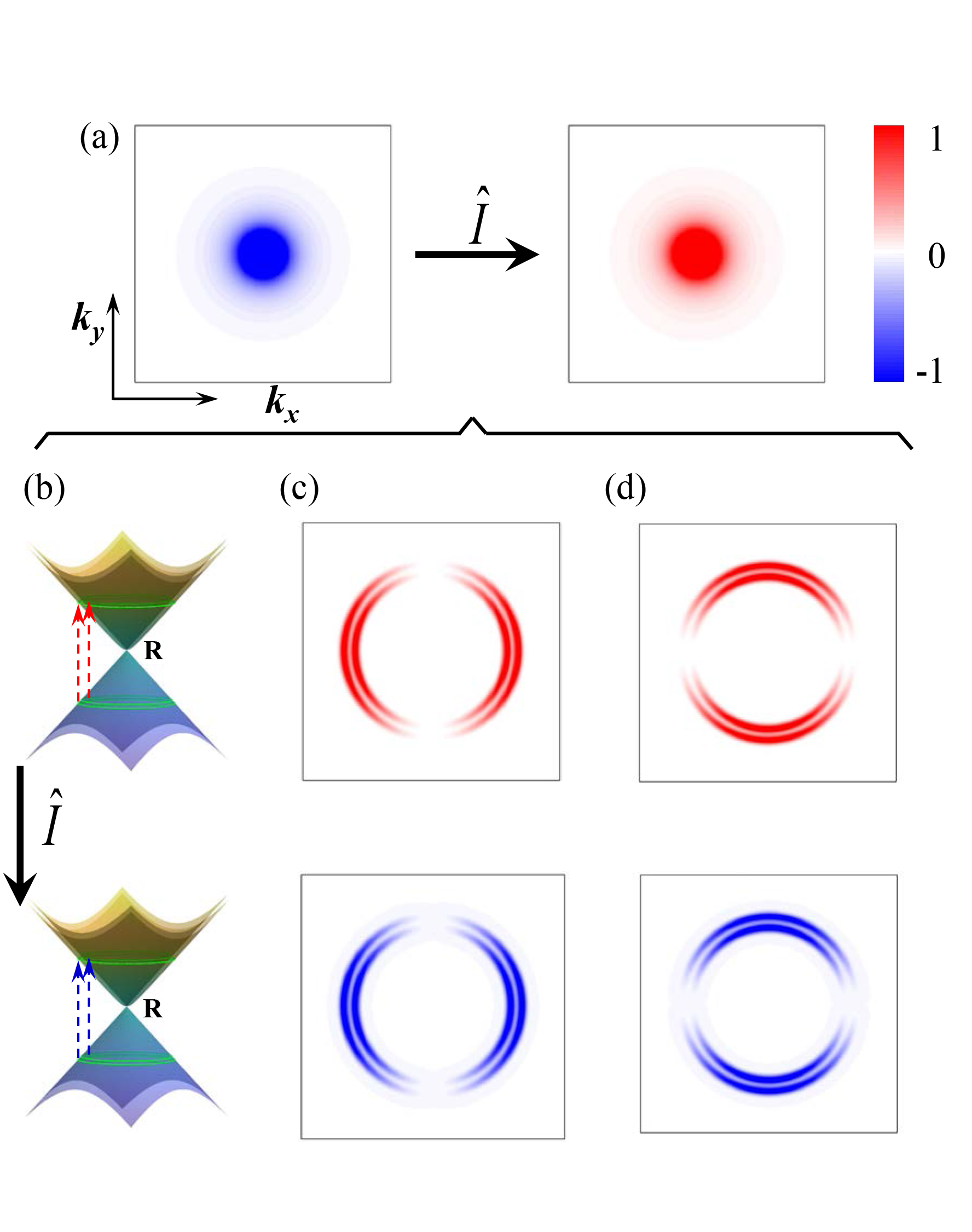}
\end{center}
\caption{
        (a) Inversion-operation-induced sign change for
        local Berry curvature ($\Omega_{z}$) distribution around
        multifold chiral fermions.
        (b) Schematic of light excitation between the lower 
        and upper cones of multifold fermions. The green rings
        represent the transition path for a given frequency.
        The light excitation from the hot ring-like
        distribution for (c)
        $\tilde{\chi}_{yz}^{CPGE,x}(\vec{k};0,\omega,-\omega)$
        and (d)
        $\tilde{\chi}_{yz}^{LPGE,x}(\vec{k};0,\omega,-\omega)$.
        $\tilde{\chi}_{yz}^{CPGE,x}$ and $\tilde{\chi}_{yz}^{LPGE,x}$
        change the sign with the inversion operation
        on the crystal structure.
        The plot is in the $k_z$=$\pi$ plane around the $R$ point.
        The color bars are in arbitrary units.
}
\label{bc_distribtution}
\end{figure}

There are mainly two types of transports based on second-order responses:
nonlinear optical effects and the nonlinear Hall
effect from Berry curvature dipole~\cite{kraut1979,belinicher1980,sipe2000,sodemann2015quantum,ma2019observation}.
The specific crystal symmetry makes the off-diagonal elements of
the nonlinear Hall conductivity tensor zero, and only one independent
diagonal element is allowed. However, owing to time-reversal
symmetry,
\begin{equation}
\begin{aligned}
	&\intop f_{0}(D_{xx}+D_{yy}+D_{zz})d\vec{k}\\
	&=\intop f_{0}(\frac{\partial\varOmega_{x}}{\partial k_{x}}+\frac{\partial\varOmega_{y}}{\partial k_{y}}+\frac{\partial\varOmega_{z}}{\partial k_{z}})d\vec{k}\\
	&=\intop\varOmega_{x}\frac{\partial f_{0}}{\partial k_{x}}+\varOmega_{y}\frac{\partial f_{0}}{\partial k_{y}}+\varOmega_{z}\frac{\partial f_{0}}{\partial k_{z}})d\vec{k}\\
	&=\intop\overrightarrow{\varOmega}\frac{\partial f_{0}}{\partial\vec{k}}d\vec{k}\\
	&=\intop\overrightarrow{\varOmega}\delta(E-E_{F})d\vec{k}\\
	&=0
\end{aligned}
\label{BCD_zero}
\end{equation}
makes the trace of Berry curvature dipole zero; consequently, nonlinear Hall effects from 
Berry curvature dipople vanish
in materials belonging to the space group $P2_13$.
Therefore, we will take the second-order optical transports,
CPGE and the linear photogalvanic effect (LPGE),
to investigate responses to the reversal of chirality of
multifold fermions.

In crystals without an inversion
center, polarized light can generate a
photocurrent in the material~\cite{kraut1979,belinicher1980,kristoffel1980,von1981,kristoffel1982,presting1982,aversa1995,sipe2000,yao2005,ma2017direct,xu2018electrically,morimoto2018current}.
In contrast to
the photovoltaic effect in p-n junctions, the 
photovoltaic effect induced by polarized light is dependent only on the bulk
band structure and not limited by the band gap.
Therefore, the polarized-light-induced photoelectric effect is also called the bulk photovoltaic effect (BPVE).
Depending on the polarization of incident light, the
photocurrent can be classified into two types: an
injection current induced by circularly polarized light
$\frac{d\vec{j}_{a}}{dt}=\chi_{bc}^{C,a}(0,\omega,-\omega)E_{b}(\omega)E_{c}(-\omega)$
and a shift current induced by linearly polarized light
$\vec{j}_{a}=\chi_{bc}^{L,a}(0,\omega,-\omega)E_{b}(\omega)E_{c}(-\omega)$~\cite{sipe2000},
where $E_{i(j,k)}(\omega)$ is the electrical field with $i,j,k=x,y,z$.

First, we check the response of the local $k$-space 
distribution of these two nonlinear optical conductivities to the inversion of
chirality. The CPGE tensor is purely imaginary and can be 
written as~\cite{sipe2000} 
\begin{equation}
\begin{aligned}
        &\chi_{bc}^{C,a}(0,\omega,-\omega)= \\
        &\frac{e^{3}\pi}{\hbar V}\underset{\vec{k}}{\boldsymbol{\sum}}\underset{m,n}{\sum}f_{nm}^{\vec{k}}\varDelta_{\vec{k},mn}^{a}[r_{\vec{k},mn}^{c},r_{\vec{k},nm}^{b}]\delta(\hbar\omega-E_{\vec{k},mn}),
\end{aligned}
\label{shift_current}
\end{equation}
 and the LPGE tensor is real and can be written as ~\cite{sipe2000}
\begin{equation}
\begin{aligned}
        &\chi_{bc}^{L,a}(0,\omega,-\omega)= \\
        &\frac{ie^{3}\pi}{\hbar V}\underset{\vec{k}}{\boldsymbol{\sum}}\underset{m,n}{\sum}f_{nm}^{\vec{k}}(r_{\vec{k},mn}^{b}r_{\vec{k},nm}^{c;a}+r_{\vec{k},mn}^{c}r_{\vec{k},nm}^{b;a})\delta(\hbar\omega-E_{\vec{k},mn}),
\end{aligned}
\label{shift_current}
\end{equation}
with
\begin{equation}
\begin{aligned}
r_{\vec{k},nm}^{a}=\frac{v_{\vec{k},nm}^{a}}{i\omega_{\vec{k},nm}}
\end{aligned}
\label{shift_current}
\end{equation}
and
\begin{equation}
\begin{aligned}
        &r_{\vec{k},nm}^{a;b}= \\
        &\frac{i}{\omega_{\vec{k},nm}}[v_{\vec{k},nm}^{a}\varDelta_{\vec{k},nm}^{a}+v_{\vec{k},nm}^{b}\varDelta_{\vec{k},nm}^{b}-\omega_{\vec{k},nm}^{ab}\\ 
        &+\underset{p\neq n,m}{\sum}(\frac{v_{\vec{k},np}^{a}v_{\vec{k},pm}^{b}}{\omega_{\vec{k},pm}}-\frac{v_{\vec{k},np}^{b}v_{\vec{k},pm}^{a}}{\omega_{\vec{k},np}})],
\end{aligned}
\label{shift_current}
\end{equation}
where $f_{\vec{k},nm}^{\vec{k}}=f_{n}^{\vec{k}}-f_{m}^{\vec{k}}$
is the difference of the Fermi--Dirac distribution between two bands,
$E_{\vec{k},mn}=E_{\vec{k},n}-E_{\vec{k},m}$ is the energy difference between the bands,
$v_{\vec{k},nm}^{a}=\frac{1}{\hbar}<n|\partial_{a}\hat{H}|m>$
is the velocity matrix, $\varDelta_{\vec{k},nm}^{a}=v_{\vec{k},nn}^{a}-v_{\vec{k},mm}^{a}$ is the Fermi velocity difference between the bands,
and $\omega_{\vec{k},nm}^{ab}=\frac{1}{\hbar^{2}}<n(\vec{k})|\partial_{ab}^{2}H|m(\vec{k})>$.
For the convenience of analysis of the microscopic relationship between the band
structure and second-order conductivity, we set
\begin{equation}
\begin{aligned}
        &\tilde{\chi}_{bc}^{C,a}(\vec{k};0,\omega,-\omega)= \\
        &\underset{m,n}{\sum}f_{nm}^{\vec{k}}\varDelta_{\vec{k},mn}^{a}[r_{\vec{k},mn}^{c},r_{\vec{k},nm}^{b}]\delta(\hbar\omega-E_{\vec{k},mn})
\end{aligned}
\label{shift_current}
\end{equation}
and
\begin{equation}
\begin{aligned}
        &\tilde{\chi}_{bc}^{L,a}(\vec{k};0,\omega,-\omega)= \\
        &\underset{m,n}{\sum}f_{nm}^{\vec{k}}(r_{\vec{k},mn}^{b}r_{\vec{k},nm}^{c;a}+r_{\vec{k},mn}^{c}r_{\vec{k},nm}^{b;a})\delta(\hbar\omega-E_{\vec{k},mn}).
\end{aligned}
\label{shift_current}
\end{equation}

For the second-order response, $\vec{j}_{a}=\chi_{bc}^{a}E_{b}E_{c}$,
the sign of the conductivity $\chi$
will become negative on inverting each of the coordinate indices.
For the specific space group $P2_13$, the glide $c_2$ rotation symmetry
$s_{2z}$ changes the sign of the optical conductivity tensor
with the $z$-index appearing an even number of times, making them zero.
Similarly, 
when the $x$- or $y$-index appears an even number of times, the elements with the index vanish because of $s_{2x}$ or $s_{2y}$, respectively.
Furthermore, owing to the $D_2$ subgroup,
the three indexes yield the same optical conductivity,
and only one nonzero independent tensor element of each index remains for both
second-order CPGE and LPGE: 
$\chi_{yz}^{C(L),x}$=$\chi_{zx}^{C(L),y}$=$\chi_{xy}^{C(L),z}$.
Thus, taking the six-fold degeneracy at the $R$ point as an example,
we analyze the changes of distribution functions of
$\tilde{\chi}_{yz}^{C(L),x}(\vec{k};0,\omega,-\omega)$
with respect to the inversion operation.

As discussed above, the chirality of the 
chiral fermions change the sign of counterpart 
crystal structures with opposite chirality. Accordingly,
since the chiral fermions are Berry curvature monopoles,
the Berry curvatures at a given $k$-point around the six-fold
degenerate points are also opposite in two lattice crystal 
structures with inverted chiralities, as shown in
in Fig. 2(a) for $\Omega_{z}$.

The chirality of the Berry curvature is also related 
to the correlation between the band structure and polarized light.
As shown in Fig. 2(b) and (c), when the Fermi level
is at the linear crossing point, two transitions exist for
a selected transition energy in the $k_z$= 0 plan. 
These two transitions form two hot rings
for both distribution functions, 
$\tilde{\chi}_{yz}^{C,x}(\vec{k};0,\omega,-\omega)$ 
and $\tilde{\chi}_{yz}^{L,x}(\vec{k};0,\omega,-\omega)$,
as shown in the upper panels of Fig. 2(c) and (d). 
The positions of the hot rings remain the same after the inversion
operation, but their signs
switch, as shown in the bottom panels of Fig. 2(c) and (d).

\begin{figure*}[htbp]
\begin{center}
\includegraphics[width=0.95\textwidth]{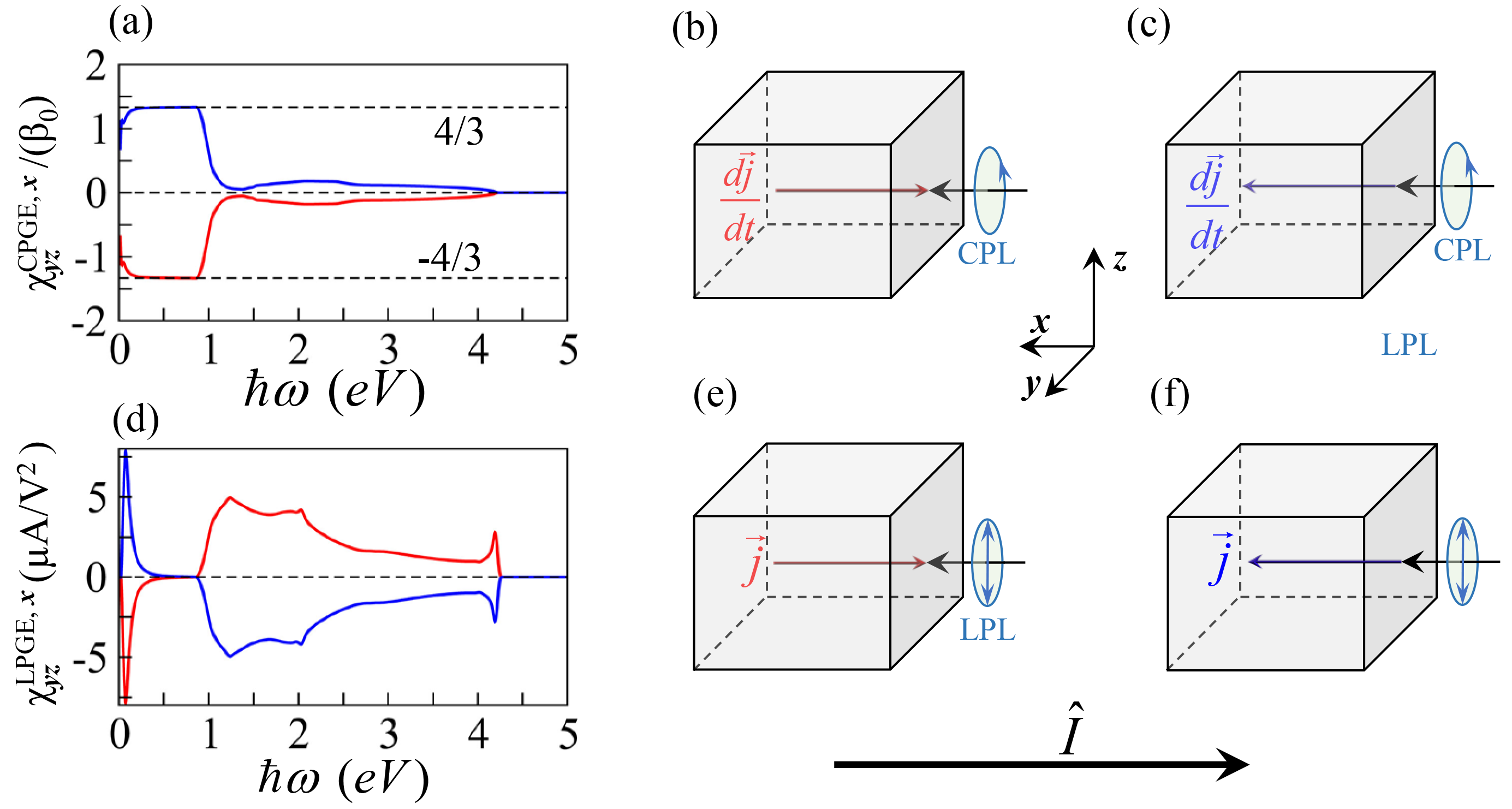}
\end{center}
\caption{
        Frequency-dependent second-order
        conductivity for (a) CPGE and (d)LPGE
	The red and blue curves correspond to
	opposite chiralities of the chiral fermions. 
	The signs of all three conductivities
        change with the switch of chirality
        of the crystal structure.
        Schematic of the experimental setup for
        identifying multifold fermions in
        anti-crystals by (b, c) CPGE and (e, f) LPGE.
        CPL in (b, c) represents circularly polarized light,
        and LPL in (e, f) represents linearly polarized light.
        The calculations were performed using the tight-binding
        model reported in Ref.~\cite{chang2017unconventional}
        with the parameters $v_1$= 1.0, $v_p$= -1.0, $v_{r1}$= 0.0,
        $v_{r2}$= -0.01, $v_{r3}$= 0.01, $v_{s1}$= -0.01, $v_{s2}$= 0.0, and
        $v_{s3}$= 0 and an onsite term $v_0$=0.01 to shift the degeneracy
        at the $R$ point to the Fermi level. $\beta_{0}=i\pi\frac{e^{3}}{h^{2}}$
	in (a).
}
\label{current}
\end{figure*}

To check the second-order optical  
conductivity, the distribution
functions must be integrated over the 
entire BZ. By employing the 
Hamiltonian of the tight-binding model~\cite{chang2017unconventional}, 
we calculated the second-order 
conductivity tensor, which fully is consistent with the
symmetry constrained tensor shapes. Owing to
the large separation of the opposite-chirality
fermions in energy space, the trace of CPGE
is a quantized value in units
$i\pi\frac{e^{3}}{h^{2}}$. In Fig. 3(a), we
show the transition-energy-dependent CPGE for
${\chi}_{yz}^{C,x}$, corresponding to
$\beta_{xx}$ in Ref.~\cite{de2017quantized}.
Since we only take one
component of the trace term, a third of the
quantized value exists in the range of $\sim$0.1
to $\sim$0.8 eV. On inverting the crystal structure
to its counterpart with opposite chirality,
the sign of CPGE changes. Therefore,
for circularly polarized light with a specific
helix, the generated voltage drop is opposite
for the same type of chiral fermions with left- 
and right-handed chirality. 

We present a schematic setup for
the experimental measurement of ${\chi}_{yz}^{C,x}$ as an example. On applying
circularly polarized light along the $x$ direction,
with its electrical field locked in the $y-z$ plane,
an injection current along $-x$ is generated.
When using the same circularly polarized light
along the same direction for the same compound with
opposite chirality, the sign of the injection current
changes, as shown in Fig. 3(b)
and (c).
Though the LPGE is not quantized, a strong shift current
is expected in this class of chiral crystals
because of the strong inversion-symmetry breaking~\cite{zhang2019strong}.
By applying linearly polarized light in the same direction,
a shift current along $x$ is generated, and
a sign change occurs for its anti-crystal, as shown in
Fig. 3(d-f). The sign change of the strong signal of
shift current can be easily detected.

 In summary, the switch of chirality of chiral 
 multifold fermions is related to the sign 
 change of charge currents generated by
 second-order electrical and optical
 responses. This relationship
 provides an effective bulk-transport approach to 
 experimentally identify the chirality of chiral 
 fermions of the same type, as well as the relationship between the chirality 
 of chiral fermions in reciprocal space 
 and the crystal structure in real space.

\begin{acknowledgments}
Thanks to Inti Sodemann for the helpful discussion. 
This work was financially supported by the ERC Advanced Grant No.\ 291472 `Idea Heusler', ERC Advanced Grant No.\ 742068 `TOPMAT'.
This work was performed in part at the Center for Nanoscale Systems (CNS), a member of the National Nanotechnology Coordinated Infrastructure Network (NNCI), which is supported by the National Science Foundation under NSF award no. 1541959. CNS is part of Harvard University.
Some of our calculations were carried out on
the Cobra cluster of MPCDF, Max Planck society.

\end{acknowledgments}

\bibliographystyle{ieeetr}
\bibliography{topology}

\end{document}